\documentclass[a4paper,11pt]{article}
\usepackage{pos}

\title{Cracks in the Standard Cosmological Model: Anomalies, Tensions, and Hints of New Physics}
\ShortTitle{Cracks in the Standard Cosmological Model}

\author*[a]{Eleonora Di Valentino}

\affiliation[a]{School of Mathematical and Physical Sciences, \\
University of Sheffield,
Hounsfield Road, Sheffield S3 7RH, United Kingdom}

\emailAdd{e.divalentino@sheffield.ac.uk}

\abstract{Cosmology has entered an era of unprecedented precision, yet increasing accuracy has revealed cracks in the standard $\Lambda$CDM paradigm. Although the model remains highly successful when confronted with individual datasets, joint analyses expose a network of tensions involving the Hubble constant, CMB lensing, curvature, neutrino masses, and the nature of dark energy.
In this contribution to the 3rd General Meeting of the COST Action COSMIC WISPers (CA21106), within the context of Working Group~2, we critically assess these discrepancies, emphasizing the role of model assumptions, parameter degeneracies, and dataset consistency. We review proposed early- and late-time solutions, discuss how recent DESI BAO results alter the viability of late-time extensions, and explore interacting dark-sector scenarios.
Our analysis highlights the need for caution in interpreting cosmological measurements and underscores the importance of internal consistency among cosmological probes before claiming percent-level accuracy or invoking new physics.}

\FullConference{3rd General Meeting of the COST Action COSMIC WISPers (CA21106) - 3rd Training School of the COST Action COSMIC WISPers (CA21106) (COSMICWISPers2025)\\
9–12 Sept 2025 and 16–19 Sept 2025\\
Sofia, Bulgaria and Annecy, France\\}

\begin{document}
\maketitle

\section{Introduction}

The standard cosmological model, commonly referred to as the $\Lambda$ cold dark matter ($\Lambda$CDM) scenario, has been widely adopted owing to its remarkable simplicity and its ability to accurately describe a broad range of astrophysical and cosmological observations~\cite{Planck:2018vyg,ACT:2025fju,SPT-3G:2025bzu,eBOSS:2020fvk,Brout:2022vxf}. Despite its empirical success, however, $\Lambda$CDM is built upon three fundamental ingredients whose physical nature remains essentially unknown: an early phase of accelerated expansion (inflation), a pressureless and collisionless dark matter (DM) component, and a cosmological constant $\Lambda$ driving the present-day accelerated expansion of the Universe.
In recent years, a growing number of observational tensions have emerged within this framework, revealing potential cracks in the standard cosmological model~\cite{Abdalla:2022yfr,DiValentino:2022fjm,CosmoVerse:2025txj}. Dark matter, and more generally the dark sector, appears to lie at the heart of many of these tensions. No longer merely the background of cosmology, the dark sector may instead provide a window onto new physics. Persistent and increasingly significant discrepancies could represent the first indirect hints of new light particles, additional relativistic species, or non-gravitational interactions beyond the Standard Model of particle physics.
At the same time, modern cosmology now probes couplings, relics, and interactions that are inaccessible to laboratory experiments, placing it in a unique position to explore fundamental physics at energies and epochs otherwise out of reach. This opportunity, however, comes with a crucial caveat. If the observed tensions originate from unresolved systematics or from incorrect assumptions underlying the $\Lambda$CDM framework itself, then any particle-physics interpretation must be approached with caution.

For the sake of simplicity, $\Lambda$CDM adopts very specific realizations of its three pillars. Inflation is typically modeled as a single, minimally coupled, slow-rolling scalar field; DM is treated as a cold, pressureless, and collisionless fluid; and dark energy (DE) is represented by a cosmological constant. Despite the theoretical shortcomings and the lack of direct physical evidence for these ingredients, $\Lambda$CDM remains the preferred framework precisely because of its ability to reproduce the observed phenomenology across a wide range of cosmological datasets.
Indeed, a flat $\Lambda$CDM model is in broad agreement with most current observations. Recent measurements from cosmic microwave background (CMB) experiments such as Planck~\cite{Planck:2018nkj}, SPT-3G~\cite{SPT-3G:2025bzu}, and ACT~\cite{ACT:2025fju}, baryon acoustic oscillation (BAO) data from SDSS~\cite{eBOSS:2020yzd} or DESI~\cite{DESI:2025zgx}, weak-lensing surveys including KiDS-1000~\cite{Wright:2025xka}, DESY3~\cite{DES:2021wwk,DarkEnergySurvey:2025bkf}, and HSCY3~\cite{Dalal:2023olq}, as well as Type~Ia supernova (SNIa) compilations such as DESY5~\cite{DES:2025sig}, Pantheon+~\cite{Scolnic:2021amr} and Union3~\cite{Rubin:2023ovl}, all indicate that $\Lambda$CDM provides an excellent fit when these probes are considered individually. 

This raises a key question: \textit{what does it actually mean for $\Lambda$CDM to ``agree'' with each probe?}
Within a Bayesian framework, agreement with a dataset is not, by itself, a meaningful statement. Cosmological inference proceeds by assuming a model \emph{a priori} and using the data to infer its parameter values and goodness of fit. In this sense, any model can be said to ``agree'' with a given dataset at some statistical level. The notion of agreement therefore does not quantify how informative or decisive the data are in assessing the validity of a model.
Whether a model is actually favored must instead be evaluated according to two distinct criteria. First, the model must provide a genuinely good fit to the data, as quantified by standard goodness-of-fit statistics. Second, model comparison must be performed to determine whether extensions of the model lead to a statistically significant improvement in the fit once the increased parameter space and the Occam penalty for additional complexity are taken into account. Only when both conditions are satisfied can a model be meaningfully regarded as preferred.
Moreover, an equally important requirement is internal consistency. While $\Lambda$CDM can fit each dataset individually, the cosmological parameters inferred from different probes are often not the same. The ``preferred'' $\Lambda$CDM model differs from one dataset to another, and the resulting parameter constraints are not mutually consistent. This lack of concordance suggests that, although $\Lambda$CDM remains phenomenologically successful, it may fail to provide a single, self-consistent description of all cosmological observations simultaneously.

\section{Tensions and Disagreements in $\Lambda$CDM and Their Consequences}

While the $\Lambda$CDM model remains the simplest and most widely adopted cosmological framework, analyses performed under this assumption reveal growing tensions between different observational probes. These discrepancies typically appear at the 2-3$\sigma$ level, insufficient to claim definitive inconsistency, yet firmly within the regime that warrants careful scrutiny. Taken together, they highlight the need for a critical re-examination of the assumptions underlying $\Lambda$CDM and motivate consideration of possible extensions or new physics.
For instance, assuming $\Lambda$CDM, we observe a 2.3$\sigma$ disagreement between the DESI DR2 BAO data and the Planck CMB constraints within the $\Lambda$CDM framework (Fig.~8 of~\cite{DESI:2025zgx}). This tension has grown from about 1.9$\sigma$ in the first DESI data release~\cite{DESI:2024mwx} to a higher significance in the latest analyses.
Similarly, we find up to a 2.9$\sigma$ discrepancy in the matter density parameter $\Omega_m$ when comparing the DESY5 SNIa data to DESI BAO measurements (Fig.~10 of~\cite{DESI:2025zgx}). Moreover, a combination of SPT-3G+ACT CMB data and DESI BAO data reveals a disagreement at the 3.7$\sigma$ level (Fig.~25 of~\cite{SPT-3G:2025bzu}).
An additional tension appears when comparing new ACT CMB data~\cite{ACT:2025fju} with the updated Planck PR4 (CamSpec) likelihood~\cite{Rosenberg:2022sdy}, where the disagreement rises to around 2.6$\sigma$ (Fig.~37 of~\cite{ACT:2025fju}). In contrast, using the earlier Plik PR3 likelihood from Planck 2018~\cite{Planck:2019nip}, the tension is only at the level of 1.6$\sigma$ (Fig.~37 of~\cite{ACT:2025fju}). 

As a consequence of the emerging tensions between different datasets interpreted within the $\Lambda$CDM framework, there is a growing indication for dynamical dark energy (DDE). When BAO measurements are interpreted within phenomenological DDE parameterisations, such as the Chevallier-Polarski-Linder (CPL) model~\cite{Chevallier:2000qy,Linder:2002et}, in which the DE equation of state is allowed to vary with time [$w(a)=w_0+w_a(1-a)$], and combined with Planck~\cite{Planck:2018vyg,Planck:2018nkj}, the inferred evolution of the DE sector departs from that of a pure cosmological constant at the $\sim 3\sigma$ level~\cite{DESI:2025zgx}. The significance of this deviation increases to the $\sim 3$--$4\sigma$ level once SNIa distance measurements are included~\cite{Scolnic:2021amr,Brout:2022vxf,DES:2024hip,DES:2024jxu,DES:2024upw,Rubin:2023ovl,DES:2025sig}, reaching values as high as about $3.8\sigma$ when DESI BAO, CMB, and Union3 SNIa data are combined~\cite{DESI:2025zgx}.
Importantly, this preference for DDE is not tied to any single dataset. Even if we exclude one of the key datasets, whether SNIa, CMB, or BAO, the indication for a dynamical equation of state for DE persists~\cite{Giare:2025pzu}. In particular, when adopting the CPL parameterisation, where the DE equation of state is expanded in terms of $w_0$ (its present value) and $w_a$ (its evolution with time), we consistently find a preference for $w_0 < -1$ and $w_a < 0$. This preference is robust across various data combinations, indicating that the data tend to favor a particular quadrant in the $w_0$-$w_a$ plane. The only scenario in which this indication weakens is when we simultaneously use the older SDSS BAO data together with the Pantheon+ SNIa data. In all other tested cases, the indication for DDE remains consistent~\cite{Giare:2025pzu}.
Alongside the evidence for DDE, we also encounter indications that the equation of state of DE may cross the so-called phantom dividing line (where $w=-1$) at some scale factor. In other words, there is always a particular scale factor at which the DE equation of state is equal to $-1$.
For a given value of this scale factor, the crossing corresponds to a line in the $w_0$–$w_a$ plane with a slope of $1/(1-a_c)$, where $a_c$ is the scale factor at which $w=-1$. When we examine the constraints from DESI combined with other datasets, we find that the allowed parameter space tends to align along one of these lines~\cite{Ozulker:2025ehg}. 
This alignment suggests that the data are not only hinting at DDE, but also strongly constraining the scale factor at which this phantom crossing occurs. All of the constraint lines in the $w_0$–$w_a$ plane intersect at the point corresponding to a cosmological constant ($w_0=-1$, $w_a=0$), but the data pick out a narrow range along these lines, resulting in a well-determined scale factor for the crossing~\cite{Ozulker:2025ehg}.
In summary, the fact that the constraints align along trajectories intersecting the cosmological constant point should not be interpreted as evidence that the indication for DDE is a mere artifact of parameter correlations. On the contrary, this alignment reflects a well-defined degeneracy direction that maps directly onto a specific scale factor at which the equation of state crosses the phantom divide. The clustering of constraints along these lines therefore encodes physical information, leading to a robust determination of the phantom crossing epoch and providing a clear indication of DDE behavior, rather than a spurious preference driven by correlations alone~\cite{Ozulker:2025ehg}.
We have thus established that the evidence for DDE and the crossing of the phantom dividing line is robust across different datasets. The next step is to explore whether this conclusion depends on the chosen parameterization of the DE equation of state.
Instead of relying solely on the CPL parameterization, we consider a variety of alternative forms: the Jassal-Bagla-Padmanabhan (JBP) parameterization~\cite{Jassal:2005qc}, an exponential parameterization~\cite{Dimakis:2023oje}, a logarithmic parameterization~\cite{Efstathiou:1999tm}, and the Barboza-Alcaniz (BA) parameterization~\cite{Barboza:2008rh}. Remarkably, all of these alternative parameterizations continue to show a preference for DDE at more than 4$\sigma$ significance~\cite{Giare:2024gpk}.
When we examine the behavior of the equation of state as a function of redshift, we find that the JBP parameterization, for instance, exhibits a double crossing of the phantom divide: one crossing at a redshift of about $z_c\sim0.3-0.4$ (similar to other parameterizations) and another at a redshift of about $z_c\sim4$. However, the parameterization that is most strongly preferred by a model comparison is the BA parameterization. In this case, the equation of state resembles a quintessence-like behavior today, crosses the phantom divide around $z_c\sim0.3-0.4$, but then settles into a plateau at a negative value without becoming arbitrarily large in magnitude.
In summary, changing the parameterization does not eliminate the evidence for DDE. On the contrary, it reinforces the conclusion that the data favor a scenario in which the DE equation of state is dynamic and crosses the phantom divide, regardless of the specific functional form assumed~\cite{Giare:2024gpk} (see also~\cite{Cortes:2024lgw,Shlivko:2024llw,Luongo:2024fww,Gialamas:2024lyw,Dinda:2024kjf,Najafi:2024qzm,Wang:2024dka,Ye:2024ywg,Tada:2024znt,Carloni:2024zpl,Chan-GyungPark:2024mlx,DESI:2024kob,Bhattacharya:2024hep,Ramadan:2024kmn,Pourojaghi:2024tmw,Reboucas:2024smm,Giare:2024ocw,Chan-GyungPark:2024brx,Li:2024qus,Jiang:2024xnu,RoyChoudhury:2024wri,Li:2025cxn,Wolf:2025jlc,Shajib:2025tpd,Chaussidon:2025npr,Kessler:2025kju,Pang:2025lvh,Roy:2024kni,RoyChoudhury:2025dhe,Paliathanasis:2025cuc,Scherer:2025esj,Giare:2024oil,Liu:2025mub,Teixeira:2025czm,Santos:2025wiv,Specogna:2025guo,Sabogal:2025jbo,Cheng:2025lod,Herold:2025hkb,Cheng:2025hug,Lee:2025pzo,Ormondroyd:2025iaf,Silva:2025twg,Ishak:2025cay,Fazzari:2025lzd,Smith:2025icl,Zhang:2025lam,Cheng:2025yue}).

Another notable consequence of the tensions between different datasets is the increasingly stringent upper bound on the total neutrino mass. By combining DESI BAO data with CMB observations, we find that the total neutrino mass is constrained to be less than $\Sigma m_\nu < 0.064$~eV at the 95\% confidence level (CL)~\cite{Elbers:2025vlz}. 
This result is particularly interesting because laboratory experiments and neutrino oscillation data indicate that for the normal ordering of neutrino masses, the total mass should be at least about 0.06~eV, and for the inverted ordering, at least about 0.1~eV at the 95\% CL. The fact that cosmological data now impose such a strong upper limit means that we are starting to see a tension between cosmological constraints and terrestrial measurements.
If we include additional datasets (such as SNIa, cosmic chronometers, galaxy clusters, gamma-ray bursts, or a prior on the Hubble constant), the upper limits become even more stringent. Depending on the method of tension calculation, we find that the disagreement between cosmology and terrestrial experiments can reach around 2.5$\sigma$ for the normal ordering and 3.5$\sigma$ for the inverted ordering, with some more extreme analyses pushing this discrepancy to as high as 5$\sigma$~\cite{Jiang:2024viw}.
Another intriguing consequence emerges if we allow the effective neutrino mass to take on negative values. While physically the neutrino mass cannot be negative, allowing this in the fit shows a preference for a slightly negative total neutrino mass, peaking around $\Sigma m_\nu \sim -0.1$~eV when combining DESI BAO and CMB data~\cite{Elbers:2025vlz}. This unphysical result underscores the underlying tension and the need for a careful reassessment of the model or the data.

\section{What About the CMB?}

While much of the community’s effort has been focused on dissecting BAO and SNIa data in search of possible systematic issues, there is a certain selection bias in our approach. We often place greater trust in datasets that align well with the Planck $\Lambda$CDM results and treat those that disagree with more skepticism. While it is relatively straightforward to reanalyze BAO or SNIa data, reanalyzing a CMB experiment is a far more resource-intensive task that typically requires hundreds of people and significant collaboration. This logistical challenge means that the CMB community often has better ``advertising'' and less frequent external scrutiny compared to other probes.

Let us take a closer look at the CMB itself. From the CMB, we can extract four independent angular power spectra: the temperature auto-correlation (TT), the cross-correlation between temperature and E-mode polarization (TE), the E-mode polarization auto-correlation (EE), and the B-mode polarization (BB), if detected.
Moreover, we know that within the $\Lambda$CDM framework, we can predict the gravitational lensing of the CMB with high accuracy. This lensing effect occurs because photons traveling from the surface of last scattering to us are deflected by the intervening matter distribution. In principle, if we introduce an amplitude parameter in front of this lensing effect, which we call $A_L$~\cite{Calabrese:2008rt}, it should be exactly equal to one. Any significant deviation from $A_L = 1$ would imply either systematic errors in our measurements or a hint of new physics.
This parameter $A_L$ effectively smears out or smooths the acoustic peaks in the CMB damping tail. The fact that $A_L$ must be equal to 1 acts as a consistency check that Planck data have struggled to meet. Using the Plik PR3 likelihood~\cite{Planck:2019nip}, we find that $A_L$ is greater than one at about the 2.8$\sigma$ level, representing a notable deviation from the expected value~\cite{Planck:2018vyg}. This discrepancy improves the fit to the data by reducing the chi-squared by about 9 when considering only temperature data, and by about 10 when polarization data are included as well.
The $A_L$ anomaly reflects an excess of gravitational lensing in the CMB TT data that is not supported by the lensing reconstruction itself, and this feature directly impacts the inferred values of several cosmological parameters. One significant implication is a preference for a closed universe with $\Omega_K < 0$~\cite{Planck:2018vyg,DiValentino:2019qzk,Handley:2019tkm}. A closed universe contains more matter, leading to more lensing, and thus helps to reconcile the observed lensing excess. However, this preference for a closed geometry places the Planck constraints in tension with BAO measurements. In particular, a direct comparison with SDSS BAO data reveals a discrepancy exceeding the $3\sigma$ level~\cite{DiValentino:2019qzk,Handley:2019tkm}.
Another consequence of the $A_L$ anomaly is its strong impact on the inferred total neutrino mass. Massive neutrinos suppress the growth of structure on scales smaller than their free-streaming length, thereby reducing the amplitude of gravitational lensing. An observed excess of lensing therefore drives cosmological fits toward smaller neutrino masses. Crucially, the effect of allowing an unphysical negative neutrino mass is already present in the Planck CMB data alone (Fig~2 of~\cite{Elbers:2024sha} and Fig.~3 of~\cite{Naredo-Tuero:2024sgf}). This behavior strengthens when Planck is combined with low-redshift datasets such as SDSS (Fig.~13 of~\cite{eBOSS:2020yzd}), indicating that the preference for negative neutrino masses originates in the CMB lensing anomaly itself rather than being driven by DESI BAO data.
This $A_L$ problem is further confirmed by new CMB data from experiments like SPT-3G, which, when combined with DESI, push the evidence for $A_L \neq 1$ to about 3.5$\sigma$~\cite{SPT-3G:2025bzu}. As a result, the upper limit on the total neutrino mass becomes even tighter, dropping below $\Sigma m_\nu < 0.048$~eV at the 95\% CL~\cite{SPT-3G:2025bzu}.

Now, one might wonder about the new Planck PR4 (NPIPE) analysis with the updated CamSpec likelihood, which claims to resolve the issues seen in earlier releases and restore consistency with the $\Lambda$CDM model~\cite{Rosenberg:2022sdy}. At first glance, it might appear that these updates have solved the $A_L$ and $\Omega_K$ anomalies. However, a closer inspection of the results shows that the underlying problem in the temperature power spectrum remains. The new CamSpec likelihood still rules out a flat universe at about the same level of confidence as before when using the temperature data alone (Fig.~14 of~\cite{Rosenberg:2022sdy}). 
What has changed is that the EE polarization data now pull $A_L$ and $\Omega_K$ closer to the $\Lambda$CDM expectation. However, this comes at a cost: it introduces a shift in the angular size of the sound horizon at recombination, parameter $\theta_*$, which should be the best-measured parameter in CMB experiments. As a result, we now have an internal tension of about 2.8$\sigma$ between the temperature and polarization data on $\theta_*$, which rises to over 3$\sigma$ when $A_L$ and $\Omega_K$ are allowed to vary (Fig.~16 of~\cite{Rosenberg:2022sdy}).
In other words, while the new likelihood analysis may appear to restore agreement with $\Lambda$CDM, it does so by shifting the problem elsewhere rather than eliminating it. Moreover, the reduced chi-squared values reveal a 4.5$\sigma$ tension between the $\Lambda$CDM best fit and the combined temperature and polarization data TTTEEE, indicating that the model is no longer a fully satisfactory fit to the data themselves (Table~1 of~\cite{Rosenberg:2022sdy}).

Finally, we must consider the critical role of the optical depth $\tau$, which encodes the integrated effect of reionization on the CMB. Reionization produces a characteristic ``polarization bump'' in the large-scale E-mode polarization at very low multipoles, while $\tau$ also enters the temperature power spectrum at smaller scales through the combination $A_s e^{-2\tau}$, affecting the amplitude of the damping tail. Because the low-$\ell$ polarization signal is intrinsically weak and close to the noise level, its measurement is particularly sensitive to instrumental noise and residual foreground contamination. As a result, successive improvements in data quality and foreground cleaning have led to a substantial downward revision of the inferred value of $\tau$ from WMAP to Planck~\cite{Planck:2018vyg}.
However, if the low-$\ell$ EE data of Planck are examined more conservatively, assumed to be Gaussian distributed and independent, the statistical significance of the polarization bump appears marginal. Fitting these data with a simple constant instead of a reionization-induced polarization bump still provides an acceptable description, with a p-value of about 0.063~\cite{Giare:2023ejv}. Moreover, when the analysis is restricted to the lowest multipoles, $\ell \leq 15$, the data are fully consistent with the absence of any polarization signal, remaining within $1\sigma$ of the null hypothesis (Fig.~1 of~\cite{Giare:2023ejv}).
Since the measured value of $\tau$ lies very close to the noise level, even modest statistical fluctuations or residual foreground contamination can significantly bias its determination. Yet $\tau$ plays a pivotal role in cosmological inference. If the low-$\ell$ EE data are excluded and only Planck high-multipole measurements are considered, the previously discussed tensions within $\Lambda$CDM largely disappear. In this case, the lensing amplitude becomes consistent with $A_L = 1$, the spatial curvature is compatible with $\Omega_K = 0$, and the DE equation of state reverts to $w = -1$, provided that $\tau$ takes a value around $0.08$~\cite{Giare:2023ejv} (see also~\cite{Sailer:2025lxj}).
In summary, the value of $\tau$ is pivotal in current cosmological analyses. Its impact is not limited to the large-scale E-mode polarization, but also propagates through the damping factor $A_s e^{-2\tau}$ into constraints on other parameters, including the total neutrino mass. Therefore, when the low-$\ell$ EE data are excluded, this leads to a relaxation of the upper bound on the neutrino mass, alleviating the apparent tension with terrestrial measurements~\cite{Jhaveri:2025neg}. Given that $\tau$ is currently constrained exclusively by one experiment, its determination and its consequences must therefore be treated with particular care.

\section{The Hubble Tension}

In our community, there is a tendency to interpret observations through the lens of personal, theoretical, and historical priors. When data align with our existing beliefs, we tend to label them as ``robust.'' Conversely, when data challenge those beliefs, we often dismiss or question their reliability. This is not to say that we necessarily need new physics; rather, we may have become too precise in our interpretations and not accurate enough in our overall approach.
We are often cherry-picking datasets in our papers based on convenience. Depending on which results better support our preferred conclusions, we might choose Plik PR3 or CamSpec, Pantheon+ or DESY5, DESI or SDSS. BAO, once considered a gold standard, is now questioned when it no longer fits our narrative. This selective use of data is arbitrary and undermines scientific objectivity. 

In the midst of these debates, we are also ignoring the elephant in the room: none of these discussions about new physics or systematic uncertainties can fully explain the high value of the Hubble constant. The $H_0$ tension~\cite{Verde:2019ivm,DiValentino:2020zio,DiValentino:2021izs,Perivolaropoulos:2021jda,Schoneberg:2021qvd,Shah:2021onj,Abdalla:2022yfr,DiValentino:2022fjm,Kamionkowski:2022pkx,Hu:2023jqc,Verde:2023lmm,DiValentino:2024yew,CosmoVerse:2025txj,Ong:2025cwv} remains a significant and unresolved challenge that no amount of selective data selection can fully address.
To understand the Hubble tension~\cite{Verde:2019ivm,DiValentino:2020zio,DiValentino:2021izs,Perivolaropoulos:2021jda,Schoneberg:2021qvd,Shah:2021onj,Abdalla:2022yfr,DiValentino:2022fjm,Kamionkowski:2022pkx,Hu:2023jqc,Verde:2023lmm,DiValentino:2024yew,CosmoVerse:2025txj,Ong:2025cwv}, we must first clarify what is meant by the Hubble constant, $H_0$. The Hubble constant quantifies the present-day expansion rate of the Universe, but it can be determined in fundamentally different ways.
One approach relies on observations in the local Universe. By measuring luminosity distances and recessional velocities of nearby galaxies, one can directly infer the proportionality constant relating distance and velocity, corresponding to the modern formulation of Hubble’s law. At sufficiently low redshifts, this determination is largely model-independent and rests on geometric measurements. While additional corrections are required at higher redshifts, the underlying principle remains unchanged.
The second approach infers $H_0$ from observations of the early Universe. In this case, one assumes a cosmological model for the expansion history, most commonly $\Lambda$CDM, and uses early-time observables such as the CMB to predict the value of the expansion rate today. Conceptually, this method amounts to observing the Universe at very early times and extrapolating its evolution forward using a specific theoretical framework, effectively predicting the present-day expansion rate from a model-dependent reconstruction of cosmic history.

The difficulty arises because these two approaches yield incompatible results. Under the assumption of the $\Lambda$CDM model, the value of the Hubble constant inferred from Planck CMB observations~\cite{Planck:2018vyg} is significantly lower than the value measured locally using distance-ladder techniques~\cite{Breuval:2024lsv}. By the end of 2021, this discrepancy had already exceeded the 5$\sigma$ level~\cite{Riess:2021jrx}, establishing the Hubble tension as one of the most severe and persistent anomalies in modern cosmology.
Recent measurements have only reinforced this picture. For example, the latest CMB data from the SPT-3G experiment yield $H_0 = 67.24 \pm 0.35~\mathrm{km\,s^{-1}\,Mpc^{-1}}$ when analysed assuming $\Lambda$CDM~\cite{SPT-3G:2025bzu}. In contrast, the most recent local determinations, combined into a global distance network, find $H_0 = 73.50 \pm 0.81~\mathrm{km\,s^{-1}\,Mpc^{-1}}$~\cite{H0DN:2025lyy}. The disagreement between these two values now reaches approximately 7.1$\sigma$, indicating a profound inconsistency between indirect and direct measurements.
Examining the broader landscape of $H_0$ determinations reveals a striking pattern. All measurements that rely on early-universe information and assume the $\Lambda$CDM model, whether based on CMB data alone or on BAO measurements combined with Big Bang Nucleosynthesis, consistently favor a lower value of the Hubble constant~\cite{Planck:2018vyg,ACT:2025fju,SPT-3G:2025bzu,eBOSS:2020yzd,DESI:2025zgx}. Conversely, every direct, late-universe determination based on local distance indicators points toward a significantly higher value of $H_0$~\cite{Freedman:2020dne,Birrer:2020tax,Anderson:2023aga,Scolnic:2023mrv,Jones:2022mvo,Anand:2021sum,Freedman:2021ahq,Uddin:2023iob,Huang:2023frr,Li:2024yoe,Pesce:2020xfe,Kourkchi:2020iyz,Schombert:2020pxm,Blakeslee:2021rqi,deJaeger:2022lit,Murakami:2023xuy,Breuval:2024lsv,Freedman:2024eph,Riess:2024vfa,Vogl:2024bum,Scolnic:2024hbh,Said:2024pwm,Boubel:2024cqw,Scolnic:2024oth,Benisty:2025tct,Li:2025ife,Jensen:2025aai,Riess:2025chq,Newman:2025gwg,Stiskalek:2025ktq,H0DN:2025lyy,Agrawal:2025tuv,Bhardwaj:2025kbw}. This systematic separation between model-dependent early-universe inferences and direct local measurements lies at the core of the Hubble tension and highlights the challenge of reconciling these probes within a single, self-consistent cosmological framework.
In the realm of local distance ladder measurements, there are multiple approaches to determining $H_0$. 
Combining together all these local distance ladder measurements is crucial, and this is precisely the motivation behind building a local distance network~\cite{H0DN:2025lyy}. By combining the expertise of different astronomical specialties, we carefully accounted for interdependencies and aimed for a transparent, consensus-driven measurement. During an ISSI-organized workshop in 2025, we brought together the leading teams working on the distance ladder. We collectively voted on the most reliable methods to form a baseline and explored various variants to test the robustness of the combined measurement. This effort resulted in the first fully networked, covariance-aware, multi-method combination of credible local distance indicators. We achieved a baseline $H_0$ measurement with a 1.1\% uncertainty and, when considering all measurements, a precision of 0.9\%~\cite{H0DN:2025lyy}. This comprehensive approach makes it clear that the Hubble tension does not depend on any single source.

\section{Possible Solutions to the Hubble Tension}

Before the advent of the DESI results, the central challenge was that BAO and SNIa data effectively measure a combination of the sound horizon and $H_0$, with a degeneracy between these two parameters. In other words, to achieve a higher $H_0$ in line with local measurements, one would need a smaller sound horizon (Fig.~1 of~\cite{Knox:2019rjx}). Conversely, to match the Planck $\Lambda$CDM value of $H_0$, one would need a larger sound horizon. Thus, the problem presented two broad avenues~\cite{Murgia:2016ccp, Pourtsidou:2016ico, Nunes:2016dlj, Kumar:2016zpg, Kumar:2017dnp, DiValentino:2017iww, Yang:2018uae, DiValentino:2019ffd, Yang:2020uga, Lucca:2020zjb, DiValentino:2020leo, Kumar:2021eev, Nunes:2021zzi, Gariazzo:2021qtg, Bernui:2023byc, Mishra:2023ueo, vanderWesthuizen:2023hcl, Zhai:2023yny, Liu:2023kce, Hoerning:2023hks, Pan:2023mie, Castello:2023zjr, Forconi:2023hsj, Yao:2023jau, Garcia-Arroyo:2024tqq, Benisty:2024lmj, Silva:2024ift, Giare:2024ytc, Bagherian:2024obh, Sabogal:2025mkp, Scherer:2025esj, Silva:2025hxw, DiValentino:2016hlg, DiValentino:2017rcr, Dutta:2018vmq, vonMarttens:2019ixw, Simon:2024jmu, Perez:2020cwa, Akarsu:2019hmw, DiValentino:2020naf, DiValentino:2020vnx, Yang:2021flj, DiValentino:2021rjj, Heisenberg:2022lob, Giare:2023xoc, Adil:2023exv, Gomez-Valent:2023uof, Lapi:2023plb, Krolewski:2024jwj, Bousis:2024rnb, Tang:2024gtq, Jiang:2024xnu, Manoharan:2024thb, Specogna:2025guo, Lee:2025pzo, Poulin:2018cxd, Smith:2019ihp, Niedermann:2019olb, Krishnan:2020obg, Ye:2021iwa, Poulin:2021bjr, Niedermann:2021vgd, deSouza:2023sqp, Poulin:2023lkg, Cruz:2023lmn, Niedermann:2023ssr, Vagnozzi:2023nrq, Efstathiou:2023fbn, Cervantes-Cota:2023wet, Garny:2024ums, Giare:2024akf, Poulin:2024ken, Pedrotti:2024kpn, Kochappan:2024jyf, DiValentino:2019exe, Alestas:2021luu, Ruchika:2023ugh, Frion:2023xwq, Ruchika:2024ymt, Visinelli:2019qqu, Ye:2020btb, Calderon:2020hoc, Akarsu:2021fol, Sen:2021wld, DiGennaro:2022ykp, Akarsu:2022typ, Ong:2022wrs, Akarsu:2023mfb, Anchordoqui:2023woo, Akarsu:2024qsi, Halder:2024uao, Anchordoqui:2024gfa, Akarsu:2024eoo, Yadav:2024duq, Paraskevas:2024ytz, Gomez-Valent:2024tdb, Toda:2024ncp, Gomez-Valent:2024ejh, Akarsu:2025gwi, Souza:2024qwd, Soriano:2025gxd, Akarsu:2025ijk, Escamilla:2025imi, Bouhmadi-Lopez:2025ggl, Pan:2019hac, Yang:2020zuk, Mirpoorian:2024fka, Yang:2021eud, Lee:2022gzh, Greene:2023cro, Greene:2024qis, Baryakhtar:2024rky, Seto:2024cgo, Lynch:2024hzh, Ghafari:2025eql, Schoneberg:2024ynd, Mirpoorian:2025rfp, DiValentino:2017oaw, Anchordoqui:2022gmw, Pan:2023frx, Allali:2024anb, Co:2024oek, Aboubrahim:2024spa, Smith:2025uaq, Efstratiou:2025iqi, Kumar:2025obb, GarciaEscudero:2025lef, Toda:2025kcq, Smith:2025zsg}: so-called ``late-time'' solutions that modify the expansion history after recombination and ``early-time'' solutions that alter physics before recombination.

One of the late-time solutions that operates in a relatively natural way involves allowing the DE equation of state to deviate from $-1$. In models such as $w$CDM, DE modifies the expansion history at intermediate redshifts, effectively slowing the expansion relative to $\Lambda$CDM. This results in a smaller integrated distance to last scattering and, consequently, in a higher inferred value of $H_0$. Importantly, these late-time modifications leave the sound horizon unchanged, as it is determined by physics prior to recombination, and instead reconcile local and early-universe measurements by reshaping the recent expansion history.
In particular, if the DE equation of state is allowed to enter the phantom regime, $w < -1$, the Hubble tension can in principle be fully resolved, as the inferred value of $H_0$ is raised to match local measurements.
However, complications arise once BAO and SNIa data are incorporated. The best-fit model inferred from the CMB alone, which favors a phantom-like equation of state, fails to reproduce the observed shape of the distance--redshift relation at low redshifts. When BAO and SNIa measurements are included, they pull the solution back toward a cosmological constant ($w = -1$), thereby reintroducing the tension with local determinations of $H_0$~\cite{Escamilla:2023oce}.
This occurs because geometrical parameter degeneracies that render different late-time models effectively indistinguishable from the CMB perspective, as they give the same distance to the last scattering surface, are broken by low-redshift data. In particular, the angular distances inferred from a best fit of Planck+BAO deviate significantly from the phantom DE best fit of Planck alone, with discrepancies that exceed the BAO observational uncertainties on the distances (Fig.~5 of~\cite{Escamilla:2023oce}). This demonstrates that late-time modifications alone are insufficient to fully resolve the Hubble tension.

On the other hand, in early-time solution scenarios~\cite{Poulin:2018cxd, Smith:2019ihp, Niedermann:2019olb, Krishnan:2020obg, Schoneberg:2021qvd, Ye:2021iwa, Poulin:2021bjr, Niedermann:2021vgd, deSouza:2023sqp, Poulin:2023lkg, Cruz:2023lmn, Niedermann:2023ssr, Vagnozzi:2023nrq, Efstathiou:2023fbn, Cervantes-Cota:2023wet, Garny:2024ums, Giare:2024akf, Poulin:2024ken, Pedrotti:2024kpn, Kochappan:2024jyf}, the parameter correlations between the Hubble constant and the sound horizon are in the right direction, since these models can simultaneously reduce the sound horizon and increase the inferred value of the Hubble constant. This class includes scenarios with additional relativistic degrees of freedom at recombination, increasing $N_{\rm eff}$, as well as early dark energy (EDE) models~\cite{Poulin:2018cxd}. A key feature of these solutions is that their confidence contours typically surround those of $\Lambda$CDM. As a result, the apparent reduction of the Hubble tension arises primarily from a volume effect, driven by the enlargement of the allowed parameter space. Consequently, to fully align the inferred value of $H_0$ with local measurements, it is generally necessary to impose a prior on $H_0$, effectively pulling the solution toward the locally measured value.
As a representative example, EDE introduces a scalar field that becomes dynamically relevant prior to recombination, contributing a sudden increase in the energy density that reduces the sound horizon and raises the inferred value of $H_0$. The characteristic mass scale of this field is typically of order $10^{-27}$~eV, and it is often modeled as an axion-like particle. To prevent significant modifications to the late-time DM abundance, the potential is usually chosen with an exponent $n=3$ rather than $n=1$, ensuring that the EDE component rapidly redshifts away after recombination.
In practice, when no prior on the Hubble constant is applied, EDE models tend to recover an $H_0$ value close to that of $\Lambda$CDM, leaving the tension at more than 3$\sigma$. It is only when a local prior for $H_0$ is included that the constraints shift, allowing the fraction of EDE to reach a significance above 6$\sigma$ (Fig.~2 of~\cite{Poulin:2025nfb}).

In conclusion, it will be crucial to obtain an independent measurement of the sound horizon. We forecast that this will become feasible by combining gravitational-wave standard sirens, for example from LISA, with angular BAO measurements from future experiments such as the final legacy release of DESI. In doing so, we expect to achieve a precision of about 1.5\% on the sound horizon. This level of precision would allow us to distinguish between early- and late-time solutions at roughly the 4$\sigma$ level~\cite{Giare:2024syw}.

\section{The Interacting Dark Energy Case}

In this section, we explore potential solutions to the Hubble tension in light of the recent DESI BAO measurements. Earlier BAO data from SDSS were largely consistent with the $\Lambda$CDM framework, disfavoring late-time departures from a cosmological constant. The DESI results have altered this picture by introducing new late-time indications for DDE, bringing back into consideration classes of late-time solutions for the Hubble tension with distinctive phenomenological features.
In this context, interacting DM-DE models (IDE) are once again viable candidates~\cite{Salvatelli:2013wra,Kumar:2016zpg, Caprini:2016qxs, Murgia:2016ccp, Zheng:2017asg, Kumar:2017dnp, DiValentino:2017iww,Kumar:2021eev,Gao:2021xnk,Pan:2023mie,Benisty:2024lmj,Yang:2020uga,Forconi:2023hsj,Pourtsidou:2016ico,DiValentino:2020vnx,DiValentino:2020leo,Nunes:2021zzi,Yang:2018uae,Zhang:2018mlj,vonMarttens:2019ixw,Lucca:2020zjb,Xiao:2021nmk,Wang:2024vmw,Gao:2022ahg,Zhai:2023yny,Joseph:2022khn,Bernui:2023byc,Becker:2020hzj,Hoerning:2023hks,Giare:2024ytc,Mukhopadhyay:2020bml,Escamilla:2023shf,vanderWesthuizen:2023hcl,Silva:2024ift,DiValentino:2019ffd,Zhao:2022ycr,Li:2024qso,Pooya:2024wsq,Halder:2024uao,Castello:2023zjr,Yao:2023jau,Mishra:2023ueo,Nunes:2016dlj,Silva:2025hxw,Yang:2025uyv,vanderWesthuizen:2025rip,Zhang:2025dwu,Li:2025muv}. While in the standard $\Lambda$CDM scenario DM and DE interact only gravitationally, more general frameworks allow for direct energy exchange between the two components, leading to characteristic late-time signatures that can mimic DDE or phantom-crossing behavior.
IDE models that are successful in alleviating the Hubble tension are typically phenomenological in nature. In these scenarios, the standard conservation equations for DM and DE are modified by the inclusion of an interaction rate that allows for energy exchange between the two sectors. This interaction is commonly parameterized through an energy transfer rate $Q$ proportional to the DE density $\rho_x$ and the conformal Hubble rate $H$, such that $Q = \xi H \rho_x$, where $\xi$ is a dimensionless coupling parameter that controls the strength and direction of the interaction.
In this framework, the Hubble tension can be fully resolved not simply through a volume effect, but through a more direct and robust overlap of constraints~\cite{DiValentino:2017iww}. This occurs because when $\xi$ is negative, energy is transferred from the DM sector to DE, effectively reducing the present-day DM abundance. Since the CMB acoustic peak structure tightly constrains the combination $\Omega_m h^2$, a lower DM density naturally leads to a higher inferred value of $H_0$.
However, once parameter degeneracies are broken by including SDSS BAO data, the statistical significance of the dark-sector coupling is substantially reduced. In this case, the preference for interaction is only at the $\sim 1\sigma$ level, with the inferred Hubble constant settling around $H_0 \simeq 70~\mathrm{km\,s^{-1}\,Mpc^{-1}}$~\cite{Nunes:2022bhn}. This leaves a residual tension with local measurements at approximately the $2.1\sigma$ level, placing the discrepancy in a borderline regime where it is difficult to assess whether it reflects a genuine inconsistency or a statistical fluctuation.
However, the situation changes once the new DESI BAO data are included. With DESI, a preference for DM-DE interaction emerges at more than the 95\% CL, raising the inferred value of the Hubble constant to around $71~\mathrm{km\,s^{-1}\,Mpc^{-1}}$~\cite{Giare:2024smz}. This value is in good agreement with local determinations of $H_0$. Importantly, Bayesian model comparison shows that the interacting scenario provides a goodness of fit that is statistically indistinguishable from that of the standard $\Lambda$CDM model, while simultaneously offering a natural resolution to the Hubble tension.

\section{Beyond Interacting Dark Energy: Other Dark Sector Interactions}

Finally, we explore the possibility that DM may not only interact with DE but could also interact with other light species. One well-motivated scenario is elastic scattering between DM and neutrinos, mediated by a new light particle. This interaction can be parameterized by a dimensionless coupling $u_{\nu\text{-DM}}$, which is proportional to the neutrino-DM scattering cross-section. This cross-section is typically expressed in units of the Thomson scattering cross-section and depends on the mass of the DM particle.
Increasing this coupling affects the CMB temperature power spectrum by modifying the damping tail and suppressing small-scale structure formation. While Planck’s range of multipoles can detect only relatively large couplings, ground-based telescopes like ACT and SPT, which observe multipoles beyond $\ell > 3000$, open a new observational window. In this regime, even small couplings have a more pronounced impact, changing the temperature power spectrum by a few percent and making such models distinguishable at high $\ell$ (Fig.~1 of~\cite{Brax:2023rrf}).

Analyzing this model, we find that Planck data alone constrain neutrino-DM scattering only through an upper limit. For couplings smaller than about $10^{-5}$, the effects are too subtle to be detected, with corrections to the CMB power spectrum below the level of one part in $10^5$. As a result, the posterior distribution becomes flat at small coupling values, indicating that these scenarios are effectively indistinguishable from the non-interacting case (Fig.~2 of~\cite{Brax:2023rrf}).
In contrast, small-scale CMB measurements reveal a clear preference for a non-zero coupling. When ACT high-$\ell$ data are combined with BAO measurements, the preferred value is $\log_{10}(u_{\nu\text{-DM}}) \simeq -4.86$ at the 68\% CL. Importantly, ACT and Planck constraints are fully consistent and show no mutual tension, as their allowed regions overlap. For couplings below $10^{-6}$, the effect again becomes too small to be detected even by ACT, leading to a plateau in the posterior (Fig.~2 of~\cite{Brax:2023rrf}).
When Planck low-$\ell$ data are combined with ACT high-$\ell$ measurements, and further with weak lensing observations from DESY3, the preference for a non-zero neutrino-DM coupling strengthens, exceeding the $3\sigma$ level~\cite{Zu:2025lrk}. This result is consistent with the suppression of small-scale clustering inferred from weak lensing data. Cosmology therefore provides a unique window onto neutrino portals and light mediators that are inaccessible to laboratory experiments.

\section{Summary and Conclusions: Where Do We Stand?}

The $\Lambda$CDM model continues to provide an impressively good fit to individual cosmological datasets. It remains a pragmatic framework whose core ingredients (dark matter, dark energy, and inflation) are employed because they work phenomenologically, rather than because they are grounded in a complete fundamental understanding. However, when all available datasets are considered simultaneously, persistent and increasingly significant cracks emerge.
We are facing a Hubble constant tension now exceeding $7\sigma$ across multiple independent methods, a CMB lensing anomaly, hints of spatial curvature, and the determination of a low optical depth that together challenge the internal consistency of the model. At the same time, cosmological constraints on neutrino masses are becoming increasingly difficult to reconcile with terrestrial experiments, while BAO and SNIa data point toward possible dynamical behavior in the dark energy sector.

The overarching lesson is that precision cosmology is meaningful only if the underlying data are internally consistent and robust. Otherwise, there is a risk of mistaking artifacts for discoveries, turning precision into a false sense of certainty. As cosmological measurements continue to improve, it will be essential to let the data speak honestly, even when this requires re-examining long-standing assumptions and methodologies, before claiming to measure the Universe at the percent level.

\begin{acknowledgments}
EDV is supported by a Royal Society Dorothy Hodgkin Research Fellowship. This publication is based upon work from the COST Actions ``COSMIC WISPers'' (CA21106) supported by COST (European Cooperation in Science and Technology).
\end{acknowledgments}

\bibliographystyle{unsrt}
\bibliography{biblio.bib}

\end{document}